\documentclass[aps,10pt,nofootinbib,groupedaddress,preprintnumbers,superscriptaddress]{revtex4}
\usepackage[dvipdfmx]{graphicx}
\usepackage{amssymb,amsfonts,amsmath,bm,color,multirow,,hyperref,cases, empheq}
\usepackage{mathrsfs}

\usepackage{enumerate}
\usepackage{ulem}

\hypersetup{%
 setpagesize=false,
 bookmarksnumbered=true,%
 bookmarksopen=true,%
 colorlinks=true,%
 linkcolor=blue,%
 citecolor=red}

\def\a{\alpha}
\def\b{\beta}
\def\D{\Delta}
\def\m{\mu}
\def\n{\nu}
\def\pa{\partial}
\def\s{\sigma}
\def\S{\Sigma}
\def\ve{\varepsilon}
\def\vp{\varphi}
\def\={\equiv}
\def\2{I\hspace{-.1em}I}
\def\3{I\hspace{-.1em}I\hspace{-.1em}I}
\def\4{I\hspace{-.1em}V}

\begin{document}

\title{\large{Escape probability of a photon emitted near the black hole horizon}}

\author{Kota Ogasawara}
\email{kota@tap.scphys.kyoto-u.ac.jp}
\affiliation{Theoretical Astrophysics Group, Department of Physics, Kyoto University, Kyoto 606-8502, Japan}

\author{Takahisa Igata}
\email{igata@rikkyo.ac.jp}
\affiliation{Department of Physics, Rikkyo University, Tokyo 171-8501, Japan}

\author{Tomohiro Harada}
\email{harada@rikkyo.ac.jp}
\affiliation{Department of Physics, Rikkyo University, Tokyo 171-8501, Japan}

\author{Umpei Miyamoto}
\email{umpei@akita-pu.ac.jp}
\affiliation{RECCS, Akita Prefectural University, Akita 015-0055, Japan}

\preprint{RUP-19-26}

\begin{abstract}
We investigate the escape of photons from the vicinity of the horizon to infinity in the Kerr-Newmann black hole spacetime.
We assume that a light source is at rest in a locally nonrotating frame and photons are emitted isotropically.
Then, we evaluate the escape probability of the emitted photons. 
The main result of this paper is the following.
If the black hole is extremal with the nondimensional spin parameter $a_*> 1/2$, however close to the horizon the light source would be, the escape probability remains nonzero.
The near-horizon limit value of the escape probability is a monotonically increasing function of $a_*$ and takes a maximum $\sim$29.1\% at $a_*=1$, i.e., for the extremal Kerr case.
On the other hand, if the black hole is extremal with $0\leq a_*\leq 1/2$ or if the black hole is subextremal, the near-horizon limit value is zero. 
\end{abstract}

\maketitle

\section{Introduction}

Very recently, the Event Horizon Telescope collaboration has succeeded in the first observation of the black hole shadows of the core of the galaxy M87~\cite{Akiyama:2019cqa}.
Although it remains controversial whether the central object forming the shadow is a black hole or not~(see, e.g., Ref.~\cite{Cardoso:2019rvt} for a review), at least we have observed photon ring scale near the ultracompact object directly through electromagnetic waves.
As the observation progresses further, we will be able to clarify various properties of the center in the future.
The brightness around the shadow depends on how often photons can escape from the source to infinity (i.e., the escape probability). Therefore, the evaluation of the probability is an important issue closely related to shadow observation.

The escape probability is also crucial in high-energy physics. In high-energy astrophysics, energetic particles are produced through the Penrose process~\cite{Penrose:1969rnc, Piran:1975apj} in the ergoregion, 
which plays a fundamental role in high-energy phenomena of our Universe. On the other hand, in the context of dark matter searches of high-energy astroparticle physics, exotic particles are produced by arbitrarily high-energy collisions near the horizon through the Banados--Silk--West process~\cite{Banados:2009pr} (see also, e.g.,  Ref.~\cite{Harada:2014vka} for a review). More recently, energy extraction process in a head-on collision---a combined process of the BSW and the Penrose---has attracted attention because the upper limit of its energy efficiency is of order 10~\cite{Schnittman:2014zsa}.
Moreover, it was shown that arbitrarily high efficiencies are achieved in the horizon limit by allowing that at least either of the colliding particles is emitted outwardly in the vicinity of the horizon before the collision.
This is called the super Penrose process~\cite{Berti:2014lva}.    
The observability of such high-energy or exotic particles depends considerably on the escape probability.

In general, the production rate of high-energy particles in the ergoregion increases as the generating point approaches the horizon, while the escape probability of photons emitted from there decreases~\cite{Chandrasekhar:1985kt}.
Since a black hole is spacetime region where nothing can escape, we can expect that the probability becomes zero if the emission arbitrarily approaches the horizon. 
Therefore, it is most likely that the observation of such energetic particles coming from the vicinity of the horizon seems impossible. 
In a previous work, however, we showed that the photon escape probability has a nonzero limiting value when the emission point arbitrarily approaches the horizon of the extremal Kerr black hole~\cite{Ogasawara:2016yfk}. We found it in the context of the observability of high-energy photons produced by the super-Penrose process.
This result implies that distant observers can see high-energy photons emitted from the vicinity of a rapidly rotating black hole. 

Nonzero escape probability of photons in the horizon limit further suggests observability of various phenomena near the horizon.
Although clarifying the conditions for this phenomenon is of significance, we still have not fully understood them.
The extremal limit of a black hole seems to be a key to this phenomenon, but so far it is unclear whether this is a sufficient condition. 

The purpose of this paper is to clarify the conditions for the escape probability remaining nonzero value when an emission point arbitrarily approaches the horizon. To achieve this, we adopt the Kerr--Newman spacetime as the background.
Since the metric includes an electric charge parameter, the extremal limit is a one-parameter family of the black hole spin under fixed mass scale.
The spin dependence allows us to consider the contribution of the extremal black hole rotation to the nonzero escape probability.
In addition, we assume that the source stays at rest in the locally nonrotating frame (LNRF) and emits photons isotropically. 
These assumptions are motivated by the fact that the center-of-mass frame in a collisional Penrose process coincides with the LNRF. Moreover, since the LNRF does not co-rotate locally with the spacetime by definition, we can remove the contribution of proper orbital rotational motion of a source to the escape probability.

This paper is organized as follows.
In the following section, we introduce the LNRF and obtain the components of the 4-momentum of a photon in the Kerr--Newman spacetime.
In Sec.~\ref{sec:3}, we show the ranges of an impact parameter for a photon that can escape from the vicinity of the horizon to infinity.
In Sec.~\ref{sec:4}, focusing on the emission from a light source at rest in the LNRF, we introduce emission angles.
The parameter ranges obtained in the previous section lead to the solid emission angle in which a photon can escape to infinity. 
Assuming isotropic emission, we define the escape probability. 
In Sec.~\ref{sec:5}, we evaluate the escape probability and confirm that it becomes zero as the emission point approaches the horizon in general.
However, in the extremal Kerr--Newman black hole with spin larger than a specific value, we observe that the probability remains a finite value even in the horizon limit.  
Section~\ref{sec:6} is devoted to conclusion and discussions. 
In this paper, we use units in which $c=1$ and $G=1$.

\section{General geodesic motion and locally nonrotating frame in the Kerr--Newman spacetime}

The Kerr--Newman metric in the Boyer--Lindquist coordinates is given by
\begin{align}
g_{\m\n}\:\!\mathrm{d}x^\m \mathrm{d}x^\n
=-\frac{\S\D}{A}\mathrm{d}t^2+\frac{\S}{\D}\mathrm{d}r^2+\S \:\!\mathrm{d}\theta^2+\frac{A}{\S}\sin^2\theta\left(\mathrm{d}\varphi-\frac{a(r^2+a^2-\Delta)}{A} \mathrm{d}t\right)^2,
\label{metric:Kerr}
\end{align}
where
\begin{align}
\S\=r^2+a^2\cos^2\theta,~~
\D\=r^2-2Mr+a^2+e^2,~~
A\=\left(r^2+a^2\right)^2-a^2\Delta\sin^2\theta.
\end{align}
This metric is parametrized by three parameters, mass $M$, spin $a$, and charge $e$.  
The spin parameter $a$ is related to the angular momentum $J$ with respect to the rotational axis 
as $a=J/M$.
Without loss of generality, we assume that
$a\geq 0$.
Throughout this paper, 
we only consider the parameter range of the black hole spacetime, $a^2+e^2\leq M^2$.
Then the event horizon is located at the radius
$r=r_{\mathrm{H}}\=M+\sqrt{M^2-a^2-e^2}$, where $\D$ vanishes.
The spacetime is stationary and 
axisymmetric with corresponding two Killing vectors $\xi^\m$ and $\psi^\m$, where $\xi^\m\pa_\m=\pa_t$ and $\psi^\m\pa_\m=\pa_\vp$.

Let us consider null geodesic motion  with 4-momentum $k^\m$ in the Kerr--Newman black hole spacetime.
By using the Hamilton--Jacobi method~\cite{Carter:1963}, the components of 
$k^\mu$
are given by
\begin{align}
k^t&=\frac{1}{\S}\left[a(L-aE\sin^2\theta)+\frac{r^2+a^2}{\Delta}\left[(r^2+a^2)E-aL\right]\right],
\\
k^r&=\frac{\sigma_r}{\S}\sqrt{R},
\\
k^\theta&=\frac{\sigma_\theta}{\S}\sqrt{\Theta},
\\
k^\vp&=\frac{1}{\S}\left[\frac{L}{\sin^2\theta}-aE+\frac{a}{\Delta}\left[(r^2+a^2)E-aL\right]\right],
\label{def:p^mu_BL}
\end{align}
where $\sigma_r$, $\sigma_\theta=\pm$, and
\begin{align}
R&\=\left[\:\!
(r^2+a^2)E-aL
\:\!\right]^2-\Delta\left[\:\!
(L-aE)^2+{\cal Q} \:\!
\right],
\\
\Theta&\={\cal Q}-\cos^2\theta\left[\:\!\frac{L^2}{\sin^2\theta}-a^2E^2\:\!\right].
\label{PTR}
\end{align}
Here $E=-\xi^\m k_\m$, $L=\psi^\m k_\m$, and $\cal Q$ are the conserved energy, angular momentum, and Carter constant, respectively.

We introduce the LNRF~\cite{Bardeen:1972} that is a tetrad basis associated with observers who corotate with the background spacetime.
The basis one-forms are given by
\begin{align}
e^{(0)}=\sqrt{\frac{\S\D}{A}} \:\!\mathrm{d}t, ~~
e^{(1)}=\sqrt{\frac{\S}{\D}} \:\!\mathrm{d}r, ~~
e^{(2)}=\sqrt{\S} \:\!\mathrm{d}\theta, ~~
e^{(3)}=\sqrt{\frac{A}{\S}}
\sin\theta
\:\!\mathrm{d}\vp
-\frac{a(r^2+a^2-\Delta)\sin\theta}{\sqrt{\S A}} \mathrm{d}t.
\label{tetrad_LNRF}
\end{align}
These satisfy $g_{\m\n}=\eta_{(a)(b)}e^{(a)}_\m e^{(b)}_\n$, where $\eta_{(a)(b)}={\rm diag}(-1,1,1,1)$, and $a$ and $b$ run from 0 to 3.
The tetrad components of the 4-momentum,
$k^{(a)}=e^{(a)}_\m k^\m$, are given by
\begin{align}
\label{eq:p^(0)}
k^{(0)}&=\sqrt{\frac{\D}{\S A}}\left[a(L-aE\sin^2\theta)+\frac{r^2+a^2}{\Delta}\left[(r^2+a^2)E-aL\right]\right],
\\
k^{(1)}&=\sigma_r\sqrt{\frac{R}{\S \D}},
\\
k^{(2)}&=\sigma_\theta\sqrt{\frac{\Theta}{\S}},
\\
k^{(3)}&=\frac{L}{\sin\theta}\sqrt{\frac{\S}{A}}
.\label{eq:p^(3)}
\end{align}

\section{Escape of a photon to infinity}
\label{sec:3}

We consider the emission of a photon from the equatorial plane near the horizon $(r, \theta)=(r_*, \pi/2)$ to infinity.
We adopt units in which $M=1$ and assume that $a\neq 0$ in what follows.

We derive the conditions for a photon escaping to infinity.
These are determined by investigating a radial turning point, i.e., $R=0$.
Introducing the dimensionless parameters
\begin{align}
b\=\frac{L}{E}, ~~ q\=\frac{{\cal Q}}{E^2},
\end{align}
for $E>0$ and solving $R=0$ for $b$, we obtain
\begin{align}
b=b_1(r)&\=\frac{a(e^2-2r)+\sqrt{\D\left[r^4-q(\D-a^2)\right]}}{\D-a^2},
\end{align}
and
\begin{align}
b=b_2(r)&\=\frac{a(e^2-2r)-\sqrt{\D\left[r^4-q(\D-a^2)\right]}}{\D-a^2}.
\end{align}
We note that $q$ is nonnegative (i.e., $q\geq0$) because $\Theta$ must be nonnegative and $\Theta={\cal Q}$ at the emission point $(r, \theta)=(r_*, \pi/2)$.
The allowed parameter region of $b$ for photon motion is given by
\begin{align}
b\leq b_1
~~&\mathrm{for}~~
r_ {\mathrm{H}} \leq r < 1+\sqrt{1-e^2},\\
b_2\leq b \leq b_1
~~&\mathrm{for}~~
1+\sqrt{1-e^2} \leq r,
\end{align}
where $b_2$ diverges at $r=1+\sqrt{1-e^2}$.
We note that $b_2\leq b$ for $r_ {\mathrm{H}} \leq r < 1+\sqrt{1-e^2}$ is also the allowed region, but for a negative energy photon.
Because such a photon cannot escape to infinity, we will not consider this region.

\begin{figure}[t]
\centering
\includegraphics[width=16cm]{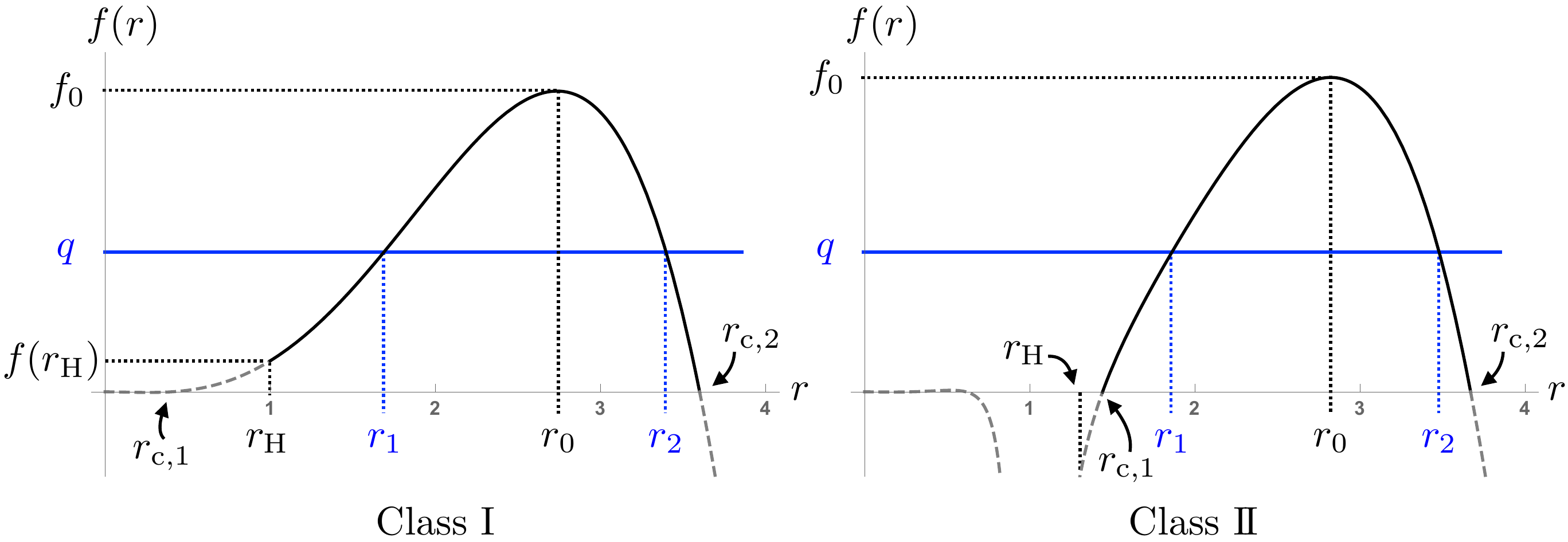}
\caption{
Typical numerical plots of $f(r)$ (black solid and gray dashed lines).
The black solid lines show $f(r)$ in the range $r>r_{\textrm{H}}$ and $0\leq f\leq f_0$.
The intersections of the blue solid lines $q$ and the black solid lines $f$ give $r_1$ and $r_2$.
We note that $r_1$ ($r_2$) is monotonically increasing (decreasing) with respect to $q$.
}
\label{fig:f}
\end{figure}

Now we focus on extremum points of $b_i$ ($i=1, 2$). 
The positions are determined by the equations
\begin{align}
b_i'(r)=0.
\end{align}
Solving these for $q$, 
we obtain the single equation
\begin{align}
\label{eq:q=f}
q=f(r)
\=\frac{r^2}{a^2}\left[\:\!
-\frac{4(1-a^2-e^2)(r-e^2)}{(r-1)^2}
+3-4e^2
-(r-1)(r-3)\:\!\right].
\end{align}
Note that the first term of $f$ vanishes 
in the extremal case~$a^2+e^2=1$. 
Outside the horizon, $f(r)$ has 
the unique local maximum with 
the value $f_0$
at $r=r_0$, where 
\begin{align}
r_0(e)&\=\frac{3+\sqrt{9-8e^2}}{2},\\
f_0(e)&\= f(r=r_0)
=\frac{r_0^4}{r_0-e^2}.
\end{align}
It is worth noting that $r_0$ and $f_0$ 
depend only on $e$ and 
monotonically decrease with respect to $e$ in the range
\begin{align}
&r_0(1)=2 \leq r_0 \leq 3 =r_0(0),\\
&f_0(1)=16 \leq f_0 \leq 27 =f_0(0),
\end{align}
and $q$ satisfies the inequality
\begin{align}
    0\leq q\leq f_0(e).
\end{align}
At the horizon, $f$ takes the value 
\begin{subequations}
\begin{empheq}[left={f(r_{\mathrm{H}})=\empheqlbrace}]{alignat=3}
&4-\frac{1}{a^2} &  &\mathrm{for}& ~~ & a^2+e^2=1, \\
&-\frac{r_{\mathrm{H}}^4}{a^2}&  ~~&\mathrm{for}& ~~ & a^2+e^2<1.
\end{empheq}
\end{subequations}
This implies that $f(r_{\mathrm{H}})>0$ only holds for $a^2+e^2=1$ and $a>1/2$.
Then, we define two classes according to the sign of $f(r_{\textrm{H}})$ as follows:
\begin{subequations}
\label{eq:class}
\begin{empheq}{alignat=2}
&\textrm{Class~I}:&&  \hspace{5mm} \displaystyle a^2+e^2=1~{\rm and}~a>\frac{1}{2},
\\[3mm]
&\textrm{Class~\2}:&&
\left\{
\begin{array}{l}
\displaystyle a^2+e^2=1~\mathrm{and}~a\leq\frac{1}{2} \\
~~~~~~~~~~\mathrm{or}
\\
\displaystyle a^2+e^2<1.
\end{array}\right.
\end{empheq}
\end{subequations}
Under this classification, 
we consider the appearance of 
the roots of Eq.~\eqref{eq:q=f} outside the horizon, i.e., 
the radii of spherical photon orbits. 
For Class~I, 
Eq.~\eqref{eq:q=f} in the range $f(r_{\textrm{H}})< q\leq f_0$ 
has two roots $r_i(q)$ 
($r_1\leq r_2$)
outside the horizon.
On the other hand, 
the equation in the range $0\leq q \leq f(r_{\textrm{H}})$
only has the largest root $r_2(q)$ outside the horizon, while 
has the next largest root $r_1(q)$ inside the horizon. 
For Class \2, 
Eq.~\eqref{eq:q=f}
in the range $0\leq q\leq f_0$
has the two roots  
$r_i(q)$ 
($r_1\leq r_2$)
outside the horizon.
In both classes,
$r_i$ ($i=1, 2$) coincide with 
$r_0$ in the case $q=f_0$. 
In particular, if $q=0$, 
the radii $r_i$ reduce to 
those of circular photon orbits 
\begin{align}
r_{\mathrm{c}, i}\= r_i(0).
\end{align}
Therefore, in other words, the criterion of the classification in Eq.~\eqref{eq:class} is whether or not the radius of the innermost spherical photon orbit reaches the horizon.
Figure~\ref{fig:f} shows typical plots of $f(r)$ and the relations among $r_{\textrm{H}}$, $r_i$, $r_{\textrm{c}, i}$, and $r_0$. 

\begin{table}[t]
\centering
\caption{
Range
of $b$ in which a photon can escape from $(r, \theta)=(r_*, \pi/2)$ to infinity.
The last column shows two pairs ($\s_r,b$) of the marginal parameter values 
with which a photon cannot escape to infinity for each cases. 
}
\begin{tabular}{lrlccc}
\hline\hline
\multicolumn{1}{c}{~Cases~} &\multicolumn{2}{c}{$q$}  & $\s_r=+$ & $\s_r=-$ & marginal pairs of ($\s_r,b$) \\ \hline
(a): $r_1< r_{\mathrm{H}}<r_*$ & $0\leq q< f(r_{\mathrm{H}})$   &(Class~I)  
&~~~$b^{\textrm{s}}_2<b\leq b_1(r_*)$~~~ & $b_1(r_{\mathrm{H}})<b<b_1(r_*)$ & ($+,b^{\textrm{s}}_2$) and ($-,b_1(r_{\mathrm{H}})$)
\\[4mm] 
\multirow{2}{*}{(b): $r_{\mathrm{H}}\leq r_1< r_*$} 
~~~~& $f(r_{\mathrm{H}})\leq q< f(r_*)$ &(Class~I) & \multirow{2}{*}{$b^{\textrm{s}}_2<b\leq b_1(r_*)$} & \multirow{2}{*}{$b^{\textrm{s}}_1<b<b_1(r_*)$} & \multirow{2}{*}{($+,b^{\textrm{s}}_2$) and ($-,b^{\textrm{s}}_1$)} \\
& $0\leq q< f(r_*)$  &(Class~\2) 
&&& \\[4mm] 
(c): $r_{\mathrm{H}}<r_*\leq r_1$ & $f(r_*)\leq q\leq f_0$ & (Classes~I \& \2) ~
& $b^{\textrm{s}}_2<b<b^{\textrm{s}}_1$ & n/a & ($+,b^{\textrm{s}}_2$) and ($+,b^{\textrm{s}}_1$) \\
\hline\hline
\end{tabular}
\label{table:classes}
\end{table}
\begin{figure}[t]
\centering
\includegraphics[width=15cm]{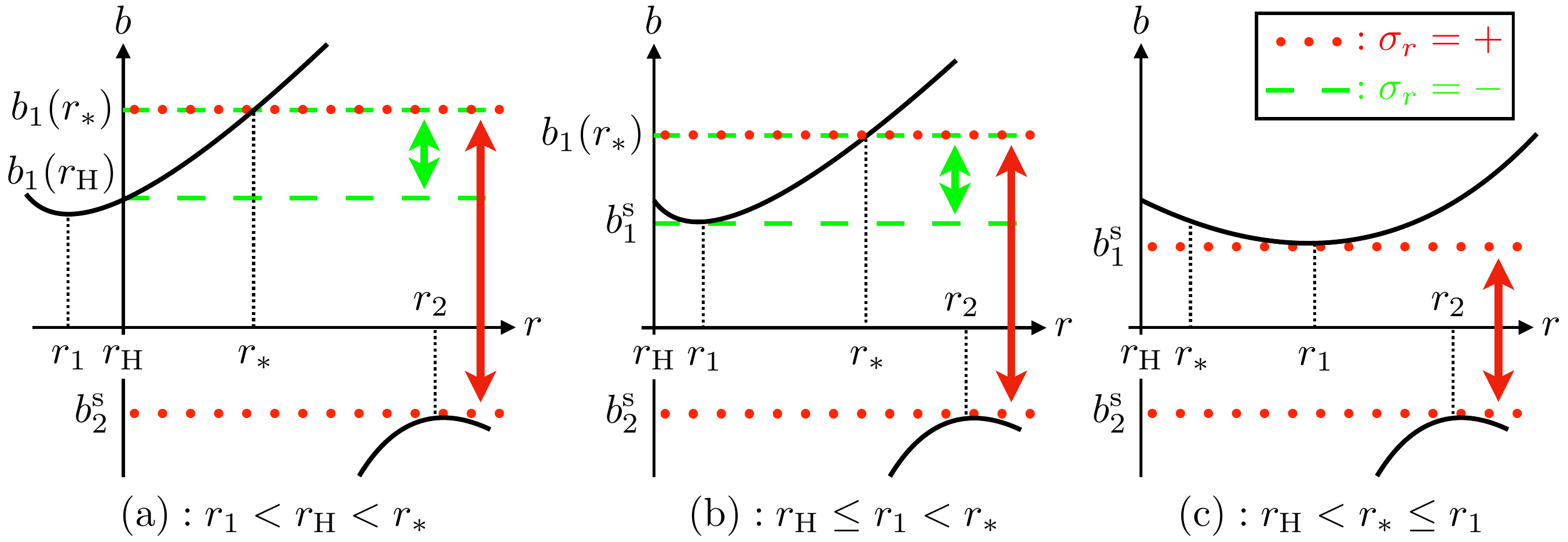}
\caption{
Schematic pictures of $b_i(r)$ in Cases~(a)--(c).
Upper and lower black solid lines denote $b_1(r)$ and $b_2(r)$, respectively.
The range of $b$ in which a photon can escape from $r=r_*$ to infinity depends on two conditions.
One is whether a photon is emitted initially radially outward ($\sigma_r=+$) or inward ($\sigma_r=-$), and the other is the
relative position of $r_1$ to $r_{\mathrm{H}}$ and $r_*$.
If photons are emitted radially outward, the maximum and minimum values of $b$ are given by the red dotted lines.
If photons are emitted radially inward, the maximum and minimum values of $b$ are given by the green dashed lines.
We note that $b_2(r)$ in the range $r<1+\sqrt{1-e^2}$ is not plotted.
}
\label{fig:b3cases}
\end{figure}

The extremum values of $b_i$ become
\begin{align}
b^{\textrm{s}}_i \= b_i(r_i)|_{q=f(r_i)}=
\frac{2(1-a^2-e^2)}{a(r_i-1)}-\frac{(r_i-1)^2}{a}+\frac{3-a^2-2e^2}{a},
\label{def:bis}
\end{align}
which are values of the impact parameter of photons on spherical photon orbits.

Let us consider the behavior of $b_i(r)$ to determine the range of $b$ in which a photon can escape from $r=r_*$ to infinity. 
From now on, we consider the case where $r_*$ is in the range $r_{\mathrm{H}} < r_* \leq r_0$. 
We define three cases according to the relative position of $r_1$ to $r_{\textrm{H}}$ and $r_*$:
\begin{subequations}
\begin{empheq}{alignat=2}
 &\textrm{Case~(a):}& ~ &r_1< r_{\mathrm{H}}<r_*, \\[1mm]
 &\textrm{Case~(b):}& ~ &r_{\mathrm{H}}\leq r_1<r_*, \\[1mm]
 &\textrm{Case~(c):}& ~ &r_{\mathrm{H}}<r_*\leq r_1 ,
\end{empheq}
\end{subequations}
where Case~(a) appears only for Class~I.

For Case~(a), as $r$ increases from $r_{\mathrm{H}}$ to $\infty$,
$b_1$ begins with $b_1(r_{\mathrm{H}})$, where
\begin{subequations}
\label{b1rH}
\begin{empheq}[left={b_1(r_{\mathrm{H}})=\empheqlbrace}]{alignat=2}
    &a+\frac{1}{a} & &(a^2+e^2=1), 
    \label{b1rH-a}\\[3mm]
    &\frac{2\left(1+\sqrt{1-a^2-e^2}\right)-e^2}{a} ~~& &(a^2+e^2<1),
    \label{b1rH-b}
\end{empheq}
\end{subequations}
and monotonically increases to $\infty$.
For Cases~(b) and (c), as $r$ increases from $r_{\mathrm{H}}$ to $\infty$, $b_1$ begins with $b_1(r_{\mathrm{H}})$, monotonically decreases to a local minimum $b^{\textrm{s}}_1$ at $r=r_1$, and monotonically increases to $\infty$.
For all cases, as $r$ increases from $r_{\mathrm{H}}$ to $1+\sqrt{1-e^2}$, $b_2$ begins with $b_2(r_{\mathrm{H}})=b_1(r_{\mathrm{H}})$ and monotonically increases to $\infty$.
As $r$ increases from $1+\sqrt{1-e^2}$ to $\infty$, $b_2$ begins with $-\infty$, monotonically increases to a local maximum $b^{\textrm{s}}_2$ at $r=r_2$, and monotonically decreases to $-\infty$.
Figure \ref{fig:b3cases} shows the schematic pictures of $b_i(r)$.

Then, the range of $b$ in which a photon can escape to infinity is given as follows.
In Case~(a), which appears only for Class~I, 
if photons are emitted radially inward (i.e., $\sigma_{r}=-$), 
then only those with $b_1(r_{\mathrm{H}})<b<b_1(r_*)$ can escape to infinity~[see the band between the green dashed lines in Fig.~\ref{fig:b3cases}(a)].
If photons are emitted radially outward (i.e., $\sigma_{r}=+$), 
then only those with $b_2^{\mathrm{s}}<b\leq b_1(r_*)$ can escape to infinity~[see the band 
between the red 
dotted lines in Fig.~\ref{fig:b3cases}(a)].
In Case~(b), which appears for both classes,
if photons are emitted radially inward, 
then only those with 
$b_1^{\mathrm{s}}<b<b_1(r_*)$ can 
escape to infinity~[see the band between 
the green dashed lines in Fig.~\ref{fig:b3cases}(b)]. 
If photons 
are emitted radially outward, 
then only those with 
$b_2^{\mathrm{s}}<b\leq b_1(r_*)$
can escape to infinity~[see the band between 
the red dotted
lines in Fig.~\ref{fig:b3cases}(b)]. 
In Case~(c), which appears for both classes,
if photons are 
emitted radially inward, 
then nothing can escape to infinity. 
If photons are
emitted radially outward, 
then only those with $b_2^{\mathrm{s}}<b< b_1^{\mathrm{s}}$ can escape to infinity~[see the band between the red dotted lines in Fig.~\ref{fig:b3cases}(c)]. 
They are summarized in Table ~\ref{table:classes}.

\section{Escape cone and critical angles}
\label{sec:4}
\begin{figure}[t]
\centering
\includegraphics[width=10cm]{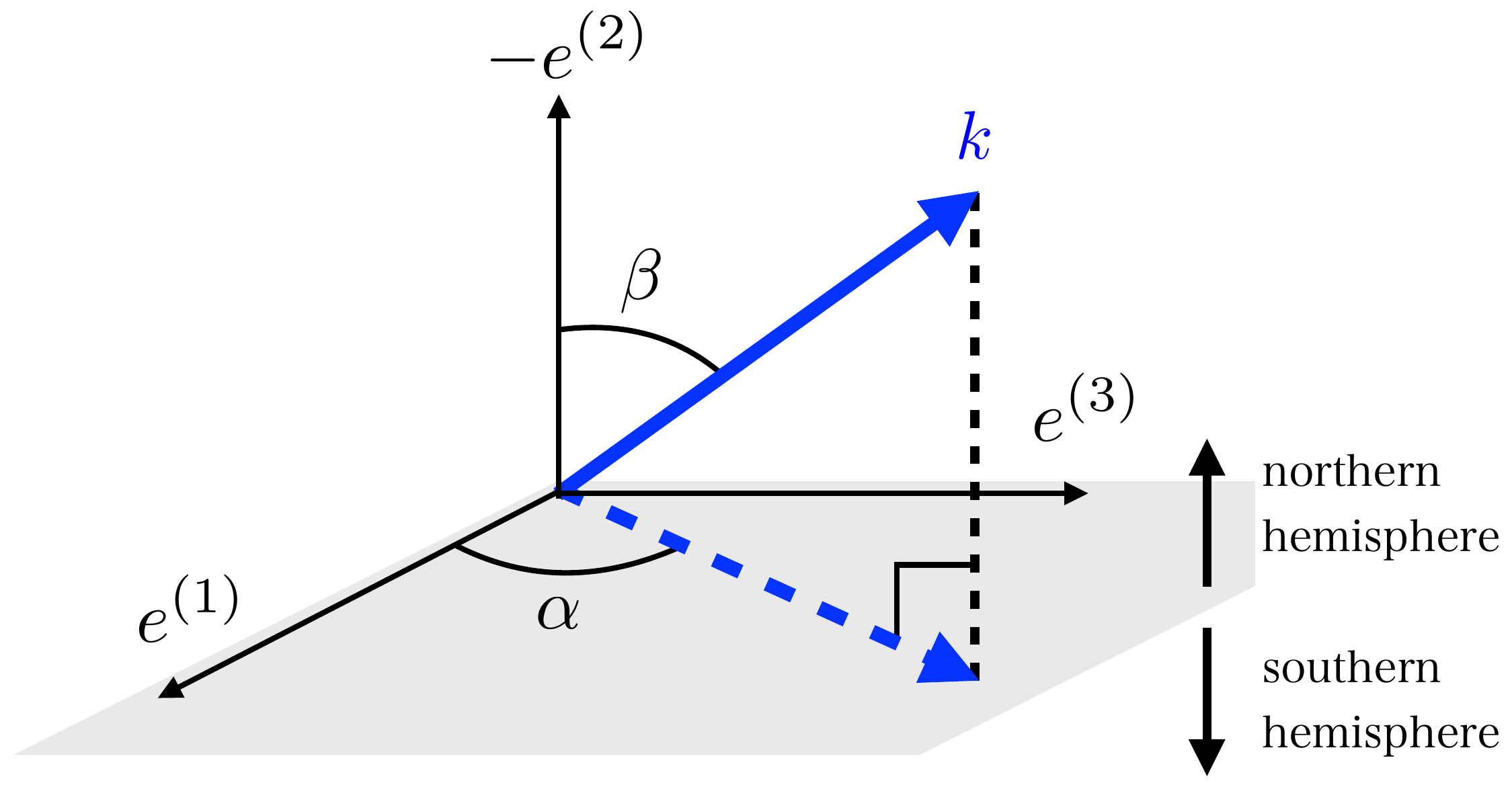}
\caption{
Emission angles $(\alpha, \beta)$ defined in the LNRF where the origin coincides with the emission point ($r,\theta$)=($r_*,\pi/2$).
Here $k$ is the projection of $k^\mu$ normal to $e^{(0)}$.
In this figure, upward and downward directions represent northward and southward directions of the Kerr--Newman black hole, respectively.
}
\label{fig:3d_angle}
\end{figure}

We introduce the photon emission angles $(\a,\b)$ 
from a light source at rest with respect to the LNRF
\begin{align}
\left.k^{(a)}
\right|_{\substack{\begin{subarray}{l}
r=r_* \\ \theta=\pi/2
\end{subarray}}}
\propto
(1,\;\cos\alpha\sin\beta,\;-\cos\beta,\;\sin\alpha\sin\beta)
\label{p_angles}
\qquad
(-\pi\leq\a<\pi \ \mathrm{and} \ 0\leq\b\leq\pi),
\end{align}
or equivalently, 
\begin{align}
\sin\a&\=\left.\frac{k^{(3)}}{\sqrt{\left(k^{(1)}\right)^2+\left(k^{(3)}\right)^2}}\right|_{\substack{
\begin{subarray}{l}
r=r_* \\ 
\theta=\pi/2
\end{subarray}
}}
=\left.\frac{b \:\!r^2\sqrt{\D}}{\sqrt{
A \hat{R}+b^2r^4\Delta
}}\right|_{r=r_*},
\\%
\cos\a&\=\left.\frac{k^{(1)}}{\sqrt{\left(k^{(1)}\right)^2+\left(k^{(3)}\right)^2}}\right|_{\substack{
\begin{subarray}{l}
r=r_* \\ 
\theta=\pi/2
\end{subarray}
}}
=\left.
\frac{\s_r\sqrt{\!A \hat{R}
}}{\sqrt{
A\hat{R}+b^2 r^4\Delta
}}\right|_{r=r_*},
\\%
\sin\b&\=\left.\frac{\sqrt{\left(k^{(1)}\right)^2+\left(k^{(3)}\right)^2}}{\sqrt{\left(k^{(1)}\right)^2+\left(k^{(2)}\right)^2+\left(k^{(3)}\right)^2}}\right|_{\substack{
\begin{subarray}{l}
r=r_* \\ 
\theta=\pi/2
\end{subarray}
}}
=\left.\frac{\sqrt{
A\hat{R}+b^2 r^4 \Delta
}}{
a(b-a)\Delta +(r^2+a^2)(r^2-ab+a^2)
}\right|_{r=r_*},
\\%
\cos\b&\=\left.\frac{-k^{(2)}}{{\sqrt{\left(k^{(1)}\right)^2+\left(k^{(2)}\right)^2+\left(k^{(3)}\right)^2}}}\right|_{\substack{
\begin{subarray}{l}
r=r_* \\ 
\theta=\pi/2
\end{subarray}
}}
=\left.-\frac{\s_\theta\sqrt{q\D A}}{
a(b-a)\Delta +(r^2+a^2)(r^2-ab+a^2)
}\right|_{r=r_*}
,\label{sincos}
\end{align}
where $\hat{R}=R/E^2$.
Thus $\alpha$ is the angle between 
$e^{(1)}$ and $k$ projected onto the equatorial plane, 
which is positive in the direction $e^{(3)}$, 
and 
$\beta$ is the angle between 
$-e^{(2)}$ and $k$, where $k$ is the projection of $k^\mu$ normal to $e^{(0)}$.
We illustrate the relation between the LNRF and 
$(\alpha, \beta)$ in Fig.~\ref{fig:3d_angle}. 
The definition of $\alpha$, $\beta$ means that the angles are functions of $\s_r$, $\s_\theta$, $b$, $q$, and $r_*$ as
\begin{align}
\a(\s_r,b,q, r_*),~~\b(\s_\theta,b,q,r_*).
\end{align}
The range of $\alpha$ is restricted depending on the sign of $b$ and $\sigma_r$ as follows.
If photons are emitted radially inward (i.e., $\sigma_r=-$), we have
\begin{alignat}{3}
-&\pi\leq\alpha\leq-\frac{\pi}{2}&\    &\mathrm{for} &~ &b\leq0,
\\[1mm]
&\frac{\pi}{2}\leq \alpha<\pi&  &\mathrm{for} &~ &b>0,
\end{alignat}
where $\alpha=-\pi$ holds for $b=0$.
If photons are emitted radially outward (i.e., $\sigma_r=+$), we have
\begin{alignat}{3}
-&\frac{\pi}{2}\leq\alpha\leq0 &\ \  &\mathrm{for} &~ &b\leq0, \\[1mm]
&0\leq \alpha\leq\frac{\pi}{2} & &\mathrm{for} & &b\geq0,
\end{alignat}
where $\alpha=0$ holds for $b=0$.
The range of $\beta$ depends on $\sigma_\theta$ as
\begin{alignat}{3}
&0\leq\beta\leq\frac{\pi}{2} & \ \  &\mathrm{for} & ~ &\sigma_\theta=-, \\[1mm]
&\frac{\pi}{2}\leq\beta\leq\pi & &\mathrm{for} & &\sigma_\theta=+.
\end{alignat}
Note that when $q=0$, $k$ is confined in the equatorial plane (i.e., $\beta=\pi/2$).

In terms of $\alpha$ and $\beta$, we define an escape cone $S$ to be the solid angle of emission that allows photons to escape to infinity.
Assuming that photons are emitted isotropically, we can define the escape probability $P(r_*)$ by
\begin{align}
P(r_*)\=\frac{1}{4\pi}\int_{S}\mathrm{d}\a \;\mathrm{d}\b\;\sin\b.
\label{def:P}
\end{align}
Since the Kerr--Newman spacetime is symmetric to the equatorial plane, 
the escape probability of isotropic emission to the northern hemisphere is 
the same with that to the southern hemisphere. 
Therefore we only need to consider the emission to the northern hemisphere (i.e., $\s_\theta=-$).
To evaluate $P$, we determine 
the critical angles, which are points on the boundary of $S$.
There is a one-to-one correspondence between 
the critical angles and 
the parameter set ($\s_r,b$, $q$) of photons that cannot marginally escape to infinity. 

Let us specify such parameter sets 
for each case. 
In Case~(a), 
if photons with $b=b_1(r_{\mathrm{H}})$ are emitted radially inward (i.e., $\sigma_r=-$), 
then they arbitrarily approach the horizon 
but cannot escape to infinity~[see Fig.~\ref{fig:b3cases}(a)].  
If photons with $b=b^{\mathrm{s}}_{2}$ 
are emitted radially outward (i.e., $\sigma_r=+$), 
then they arbitrarily approach $r=r_2$ but 
cannot escape to infinity. 
In Case~(b), 
if photons with $b=b_1^{\mathrm{s}}$ are emitted radially inward, then they arbitrarily approach $r=r_1$~[see Fig.~\ref{fig:b3cases}(b)].
If photons with $b=b_2^{\mathrm{s}}$
are emitted radially outward, 
then they arbitrarily approach $r=r_2$. 
In Case~(c), 
if photons with $b=b_i^{\mathrm{s}}$ are emitted radially outward, then they arbitrarily approach $r=r_i$ but 
cannot escape to infinity~[see Fig.~\ref{fig:b3cases}(c)].

\begin{figure}[t]
\centering
\includegraphics[width=12cm]{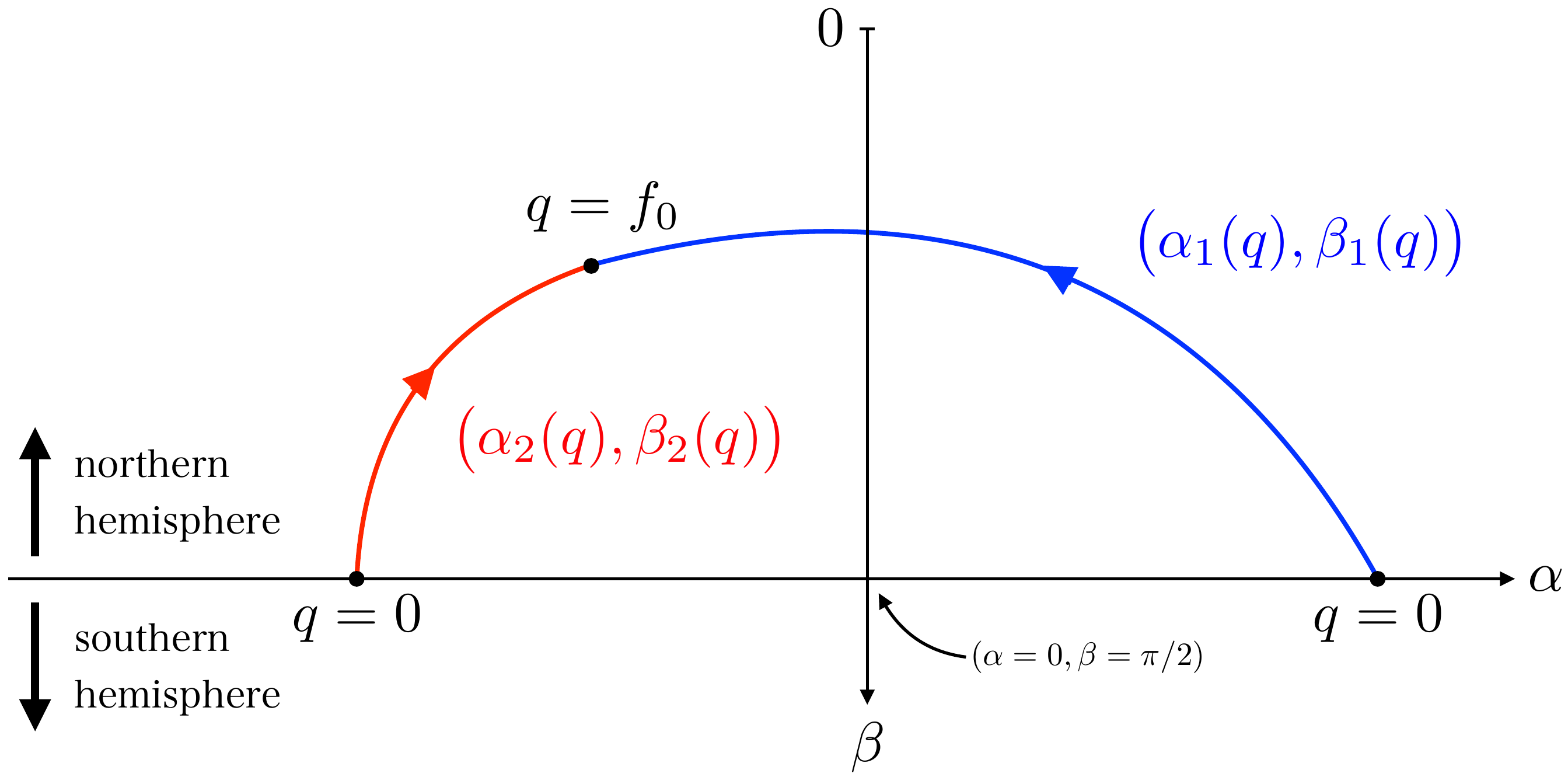}
\caption{
Schematic picture of the critical angles with fixed $r_*$ for the northern hemisphere in $\a$-$\b$ plane.
The blue and red solid lines show ($\a_1(q),\b_1(q)$) and ($\a_2(q),\b_2(q)$) for $0\leq q \leq f_0$, respectively.
We plot the critical angles as functions of $q$ because we have chosen $\s_r$, $\s_\theta$, and $b$ for the critical angles.
The critical angle $\a_1(q)$ ($\a_2(q)$) is monotonically decreasing (increasing) with respect to $q$. 
We note that the critical angles in the southern hemisphere is symmetrical about the $\a$-axis.
}
\label{fig:EC}
\end{figure}

We find that all the parameter pairs appear as the boundary of the ranges for photons that can escape to infinity~[see Table~\ref{table:classes}].
Therefore, we call them the marginal pairs. 
They are summarized in the last column of Table~\ref{table:classes}. 
Finally we obtain the critical angles $(\alpha_i, \beta_i)$ ($i=1, 2$) relevant to marginal parameter values associated with $b_i$ and their total set $\partial S$ (i.e., the boundary of $S$) as follows:
\begin{align}
\pa S=\bigcup_{i=1,2} \left\{ \big(\a_i,\b_i\big) \Big| ~0\leq q \leq f_0\right\},
\end{align}
where 
\begin{subequations}
\label{def:critical_angles1}
\begin{empheq}[left={\big(\a_1,\b_1\big)\equiv\empheqlbrace}]{alignat=3}
&\big(\a_\textrm{1(a)},\b_{1(\mathrm{a})}\big) 
\equiv \big(\a,\b \big)
\big|_{\substack{
\begin{subarray}{l}
\s_r=- \\ b=b_1(r_{\mathrm{H}})
\end{subarray}}} 
&~\textrm{for} 
& ~ 0\leq q< f(r_{\mathrm{H}})
&~&[\:\!\textrm{Case~(a)}\:\!], 
\\[3mm]
&\big(\a_\textrm{1(b)},\b_{1(\mathrm{b})}\big) 
\equiv \big(\a,\b \big)
\big|_{\substack{
\begin{subarray}{l}
\s_r=- \\ b=b_1^{\textrm{s}}
\end{subarray}
}}
&~\textrm{for} 
& ~ f(r_{\mathrm{H}})\leq q < f(r_*) 
&~~&[\:\!\textrm{Case~(b)}\:\!], 
\\[3mm]
&\big(\a_\textrm{1(c)},\b_{1(\mathrm{c})}\big) 
\equiv \big(\a,\b \big)
\big|_{\substack{
\begin{subarray}{l}
\s_r=+ \\ b=b_1^{\textrm{s}}
\end{subarray}
}} 
&~\textrm{for} 
&~ f(r_*) \leq q\leq f_0
&~&[\:\!\textrm{Case~(c)}\:\!],
\end{empheq}
\end{subequations}
in Class~I,
\begin{subequations}
\begin{empheq}[left={\big(\a_1,\b_1\big)\equiv\empheqlbrace}]{alignat=3}
&\big(\a_{\textrm{1(b)}},\b_{1(\mathrm{b})}\big) 
\equiv \big(\a,\b \big)
\big|_{\substack{
\begin{subarray}{l}
\s_r=- \\ b=b_1^{\textrm{s}} 
\end{subarray}
}} 
&~ \textrm{for} 
& ~0\leq q < f(r_*)
&~&[\:\!\textrm{Case~(b)}\:\!], 
\\[3mm]
&\big(\a_{\mathrm{1(c)}},\b_{1(\mathrm{c})}\big) 
\equiv \big(\a,\b \big)
\big|_{\substack{
\begin{subarray}{l}
\s_r=+ \\ b=b_1^{\textrm{s}} 
\end{subarray}
}} 
&~ \textrm{for} 
& ~ f(r_*) \leq q\leq f_0
&~~&[\:\!\textrm{Case~(c)}\:\!], 
\end{empheq}
\end{subequations}
in Class \2, and
\begin{align}
\big(\a_2,\b_2\big)\=\big(\a,\b\big)\big|_{\substack{
\begin{subarray}{l}
\s_r=+ \\ b=b_2^{\textrm{s}} 
\end{subarray}
}} ~~\textrm{for}~~0\leq q\leq f_0 ~~ [\:\!\textrm{Cases~(a)--(c)}\:\!],
\label{def:critical_angles2}
\end{align}
in Classes~I and \2.
Note that once we fix the value of $r_*$, then the critical angles $(\alpha_i, \beta_i)$ depend only on $q$, i.e., $\alpha_i=\alpha_i(q)$ and $\beta_i=\beta_i(q)$.
Figure~\ref{fig:EC} shows a schematic picture of the 
critical angles.

\section{Escape probability}
\label{sec:5}

In this section, we evaluate the escape probability $P$ 
for a photon in Eq.~\eqref{def:P}.
Let $\alpha_{\min}$ and $\alpha_{\max}$ be the minimum and the maximum of the critical angle,
and $\beta_{\min}(\alpha)$ and $\beta_{\max}(\alpha)$ be the minimum and the maximum of the critical angle for a given $\alpha$.
Assuming that $S$ is convex, we obtain
\begin{align}
P=\frac{1}{4\pi}\int^{\a_{\max}}_{\a_{\min}} \mathrm{d}\a \int^{\b_{\max}(\a)}_{\b_{\min}(\a)} \mathrm{d}\b \sin\b
=\frac{1}{2\pi} \int^{\a_{\max}}_{\a_{\min}} \mathrm{d}\a \cos\b_{\min}(\a),
\label{P_cal}
\end{align}
where we have used $\beta_{\max}=\pi-\beta_{\textrm{min}}$.
Using the critical angles $(\alpha_i, \beta_i)$
and 
the relation between $q$ and $r_i$ given in Eq.~\eqref{eq:q=f}~[see also Figs.~\ref{fig:f} and \ref{fig:EC}], 
we change the integration variable $\alpha$ of Eq.~\eqref{P_cal} to $r_i$ as
\begin{align}
P=P_1+P_2,
\end{align}
where
\begin{align}
P_i=\frac{(-1)^{i}}{2\pi}
\int_0^{f_0} \mathrm{d}q\:\!
\frac{\mathrm{d} \alpha_i}{\mathrm{d} q}\cos\b_i
=\frac{(-1)^{i}}{2\pi}
\int^{r_0}_{r_{\textrm{c},i}} \mathrm{d}r_i 
\frac{\mathrm{d} \a_i}{\mathrm{d} r_i}
\cos\b_i.
\end{align}
In particular, 
using the definition of the critical angles in Eqs.~\eqref{def:critical_angles1}--\eqref{def:critical_angles2}, 
we can reduce $P_1$ as
\begin{subequations}
\label{int1}
\begin{empheq}[left={P_1=\empheqlbrace}]{alignat=2}
&-\frac{1}{2\pi}
\int^{r_{\mathrm{H}}}_{r_{\mathrm{c,1}}} \mathrm{d}r_1 \frac{\mathrm{d} \a_{\mathrm{1(a)}}}{\mathrm{d} r_1} \cos\b_{\mathrm{1(a)}}
-\frac{1}{2\pi}\int^{r_*}_{r_{\mathrm{H}}} \mathrm{d}r_1 \frac{\mathrm{d} \a_{\mathrm{1(b)}}}{\mathrm{d} r_1} \cos\b_{\mathrm{1(b)}}
-\frac{1}{2\pi}\int^{r_0}_{r_*} \mathrm{d}r_1 \frac{\mathrm{d} \a_{\mathrm{1(c)}}}{\mathrm{d} r_1} \cos\b_{\mathrm{1(c)}}
&~&(\mathrm{Class~I}),\  
\\
&-\frac{1}{2\pi}\int^{r_*}_{r_{\mathrm{c,1}}} \mathrm{d}r_1 \frac{\mathrm{d} \a_{\mathrm{1(b)}}}{\mathrm{d} r_1} \cos\b_{\mathrm{1(b)}}
-\frac{1}{2\pi}\int^{r_0}_{r_*} \mathrm{d}r_1 \frac{\mathrm{d} \a_{\mathrm{1(c)}}}{\mathrm{d} r_1} \cos\b_{\mathrm{1(c)}}
&~&(\mathrm{Class~\2}),
\end{empheq}
\end{subequations}
It is useful to note that some of the integrands in the above integrals have a common form
\begin{align}
\left.
\frac{\mathrm{d} \a_{\mathrm{1(b)}}}{\mathrm{d} r_1} \cos\b_{\mathrm{1(b)}}
\right|_{r_1=x}
=\left.
\frac{\mathrm{d} \a_{\mathrm{1(c)}}}{\mathrm{d} r_1} \cos\b_{\mathrm{1(c)}}
\right|_{r_1=x}
=\left.
\frac{\mathrm{d} \a_2}{\mathrm{d} r_2} \cos\b_2
\right|_{r_2=x}
\=g(x).
\label{def:g}
\end{align}
Finally, from Eqs.~\eqref{P_cal}--\eqref{def:g}, 
$P$ is given by
\begin{subequations}
\label{P1}
\begin{empheq}[left={P=\empheqlbrace}]{alignat=2}
&-\frac{1}{2\pi}
\int^{r_{\mathrm{H}}}_{r_{\mathrm{c}, 1}} \mathrm{d}r_1 \frac{\mathrm{d} \a_{\mathrm{1(a)}}}{\mathrm{d} r_1} \cos\b_{\mathrm{1(a)}}
-\frac{1}{2\pi}\int^{r_{\mathrm{c, 2}}}_{r_{\mathrm{H}}} \mathrm{d}x\:\! g(x)
~~& ~ &(\mathrm{Class~I}), \ 
\label{eq:P1-I}
\\
&-\frac{1}{2\pi} \int^{r_{\mathrm{c, 2}}}_{r_{\mathrm{c, 1}}} \mathrm{d}x\:\! g(x)
& ~ &(\mathrm{Class~\2}).
\label{eq:P1-II}
\end{empheq}
\end{subequations}

\subsection{Extremal Kerr black hole}
We evaluate the critical angles and escape probability in the extremal Kerr black hole. 
Since $a=1$ and $e=0$ (i.e.,  Class~I), the function $f(r)$ in Eq.~\eqref{eq:q=f} reduces to 
\begin{align}
    f(r)=r^3(4-r),
\end{align}
which has the maximum value $f_0=27$ at $r_0=3$ and zeros at the radii of circular photon orbits 
\begin{align}
r_{\textrm{c},1}=0, 
\quad
r_{\textrm{c},2}=4.
\end{align}
Furthermore, the values of $b$ given in Eqs.~\eqref{def:bis} and \eqref{b1rH-a} are of the form
\begin{align}
&b_i^{\mathrm{s}}=-r_i^2+2r_i+1,
\\
&b_1(r_{\textrm{H}})=2,
\end{align}
where $r_{\mathrm{H}}=1$. 

Using these expressions, we can evaluate the critical angles $(\alpha_i, \beta_i)$. In Fig.~\ref{fig:EC_extremal_KN}, we find sets of critical angles $\partial S$ for several values of $r_*$.
As $r_*$ decreases toward $r_{\textrm{H}}=1$, the size of the escape cone $S$ decreases.
However, we should notice that $S$ still has enough size, even if $r_*$ is close enough to the horizon~[see $r_*=1.001$ case].
Now let us evaluate the escape probability $P$ in  Eq.~\eqref{eq:P1-I} with the above expressions.
In Fig.~\ref{fig:P_extremal_KN}, the solid black line shows $P$ as a function of $r_*$. We find that $P$ decreases as $r_*$ decreases toward $r_{\mathrm{H}}$ but is nonzero in the horizon limit. Indeed, the value of $P$ in the horizon limit takes 
\begin{align}
\lim_{r_*\to 1}P
=0.2916\cdots.
\end{align}
This result means that about 30 percent of photons isotropically emitted from a light source near the horizon of the extremal Kerr black hole can escape to infinity.

Here let us confirm that the result of Ref.~\cite{Ogasawara:2016yfk} can be reproduced.
Assuming that photons are emitted isotropically but confined in the equatorial plane, we can identify the escape cone with the segment $\alpha \in [\:\!\alpha_2, \alpha_{1(a)}\:\!]|_{q=0}$ on the line $\beta=\pi/2$ in Fig.~\ref{fig:EC_extremal_KN}(i).
Thus the escape probability in such a situation is given by
\begin{align}
P=\frac{\left.\left(\a_{\mathrm{1(a)}}-\a_2\right)\right|_{q=0}}{2\pi},
\end{align}
and the horizon limit of this expression is $5/12$.
This is consistent with the result of Ref.~\cite{Ogasawara:2016yfk}.

\subsection{Extremal Kerr--Newman black hole}
\begin{figure}[t]
\centering
\includegraphics[width=15cm]{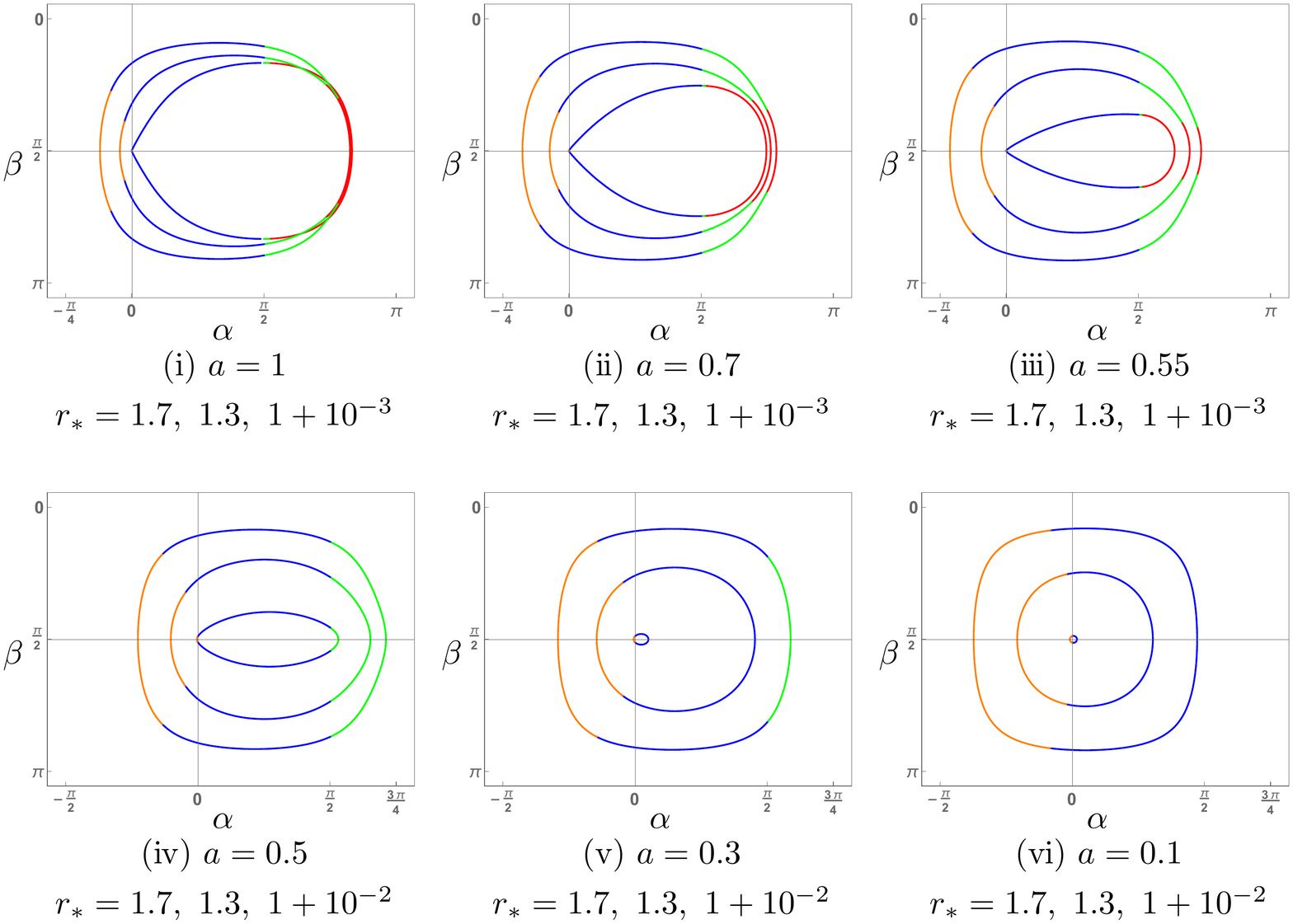}
\caption{
Critical angles in $\a$-$\b$ plane in the extremal Kerr--Newman black hole.
The red, green, blue, and orange lines show ($\a_{\rm 1(a)},\b_{\rm 1(a)}$), ($\a_{\rm 1(b)},\b_{\rm 1(b)}$), ($\a_{\mathrm{1(c)}},\b_{\mathrm{1(c)}}$), and ($\a_2,\b_2$), respectively.
We set $r_*=1.7$, $1.3$, $1+10^{-3}$ for Class~I (the upper three panels) and $r_*=1.7$, $1.3$, $1+10^{-2}$ for Class \2 (the lower three panels) from outside to inside.
The inside of each closed solid curve shows the escape cone $S$.
We can see that, in the near-horizon emission case for Class~I, the presence of the critical angles in Case~(a) prevents the escape cone from disappearing.
}
\label{fig:EC_extremal_KN}
\end{figure}

We evaluate the critical angles and escape probability in the extremal Kerr--Newman black hole (i.e., $a^2+e^2=1$). Therefore, we replace $e$ in all the above expressions with $\sqrt{1-a^2}$. 
Then, $f(r)$ reduces to
\begin{align}
f(r)&=-\frac{r^2}{a^2}(r-r_{\mathrm{c},1})(r-r_{\mathrm{c}, 2}),
\end{align}
where 
\begin{align}
r_{\mathrm{c},1}=2(1-a),
\quad
r_{\mathrm{c}, 2}=2(1+a).
\end{align}
This function has the maximum with the value
\begin{align}
f_0=\frac{(\sqrt{1+8
   a^2}-1) (\sqrt{1+8
   a^2}+3)^3}{16 a^2}
\end{align}
at $r=r_0=(\sqrt{1+8a^2}+3)/2$. 
Furthermore, the values of $b$ given in Eqs.~\eqref{def:bis} and \eqref{b1rH-a} are of the form
\begin{align}
b_i^{\mathrm{s}}=\frac{-r_i^2+2r_i+a^2}{a}, 
\quad
b_1(r_{\mathrm{H}})=a+\frac{1}{a},
\end{align}
where $r_{\mathrm{H}}=1$.

The critical angles with the above expressions are shown in Fig.~\ref{fig:EC_extremal_KN}.
We can confirm that even if $r_*$ is close enough to $r_{\mathrm{H}}$, the escape cones in Class~I have a nonzero size in the region $\alpha>0$.
On the other hand, as $r_*$ approaches $r_{\mathrm{H}}$ sufficiently, the escape cones in Class~\2 shrink to the origin of the $\alpha$-$\beta$ plane.
The critical angles $(\a_{\textrm{1(a)}},\b_{\textrm{1(a)}})$ that appears only in Class~I seem to prevent the escape cones in the horizon limit from disappearing.

\begin{figure}[t]
\centering
\includegraphics[width=10cm]{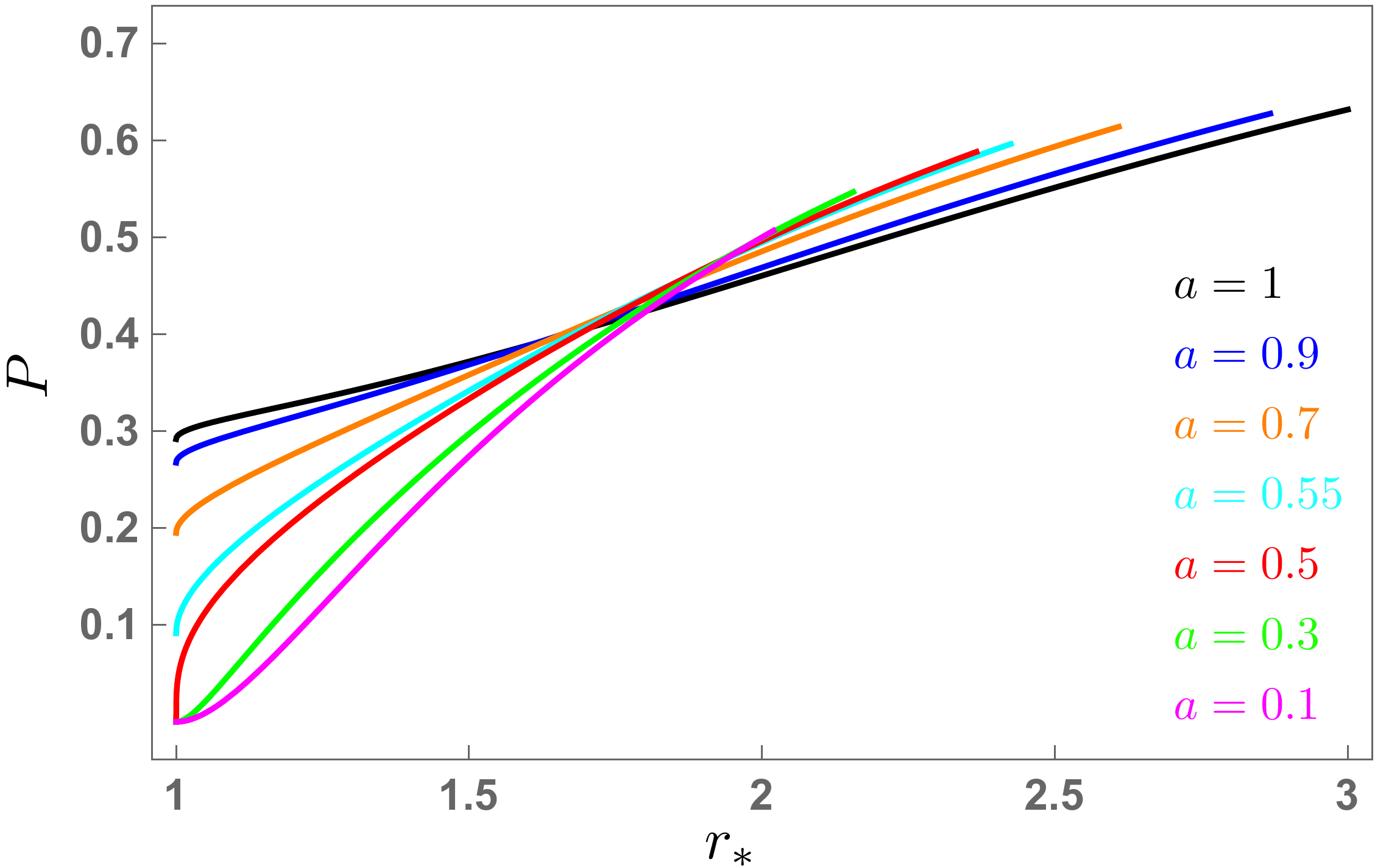}
\caption{
Escape probability in the extremal Kerr--Newman black hole.
}
\label{fig:P_extremal_KN}
\end{figure}

Figure~\ref{fig:P_extremal_KN} shows the numerical plots of the escape probability for various values of $a$.%
\footnote{
The integral of the first term in Eq.~\eqref{eq:P1-I} is evaluated analytically:
\begin{align}
-\frac{1}{2\pi}\int^{r_{\mathrm{H}}}_{r_{\mathrm{c,1}}} \mathrm{d}r_1 \frac{\mathrm{d} \a_{\rm 1(a)}}{\mathrm{d} r_1} \cos\b_{\rm 1(a)}
=&\frac{1}{2\pi}\left[
\arctan\left(\frac{(1+a^2)r^2_*}{a\left[r_*-1+(r_*+1)(r^2_*+a^2)\right]}\sqrt{\frac{4-1/a^2}{(r_*-1)(r_*-3)}}\right)
\right.\nonumber\\&\left.\hspace{10mm}
-\frac{(1+a^2)r^2_*}{a\left[r_*-1+(r_*+1)(r^2_*+a^2)\right]}
\arctan\sqrt{\frac{4-1/a^2}{(r_*-1)(r_*-3)}}
\right].
\end{align}
The horizon limit of this expression is $1/4-1/(8a)$.
} 
As $r_*$ decreases, $P$ monotonically decreases in common regardless of $a$. 
On the other hand, qualitative difference depending on $a$ appears in near-horizon emissions.
In the horizon limit, $P$ takes a nonzero value for Class~I while vanishes for Class~\2. Some explicit values of $P$ for near-horizon emissions are shown in Table~\ref{table:P_extremal_KN}. 
We should note that $P$ in the horizon limit monotonically increases as $a$ increases  in Class~I.
Hence, we find that the maximum value of $P$ in the horizon limit is the case of the extremal Kerr black hole.

\begin{table}[t]
\centering
\caption{
Escape probability 
evaluated at $r_*=1+\ve$ in the extremal Kerr--Newman black hole.
}
\begin{tabular}{llllllll}
\hline\hline
 & ~$a=1$ & ~$a=0.9$ & ~$a=0.7$ & ~$a=0.55$ & ~$a=0.5$ & ~$a=0.3$ & ~$a=0.1$ \\ \hline
 ~$\ve=10^{-1}$ & ~$3.14\times10^{-1}$ & ~$2.96\times10^{-1}$ & ~$2.45\times10^{-1}$ & ~$1.81\times10^{-1}$ & ~$1.50\times10^{-1}$ & ~$5.52\times10^{-2}$ & ~$3.02\times10^{-2}$ \\ 
~$\ve=10^{-3}$ & ~$2.93\times10^{-1}$ & ~$2.69\times10^{-1}$ & ~$1.98\times10^{-1}$ & ~$9.92\times10^{-2}$ & ~$3.27\times10^{-2}$ & ~$1.62\times10^{-5}$ & ~$4.52\times10^{-6}$ \\ 
 ~$\ve=10^{-5}$ & ~$2.91\times10^{-1}$ & ~$2.67\times10^{-1}$ & ~$1.95\times10^{-1}$ & ~$9.19\times10^{-2}$ & ~$9.91\times10^{-3}$ & ~$1.64\times10^{-9}$ & ~$4.54\times10^{-10}$ \\ 
 ~$\ve\to0$ & ~$2.91\times10^{-1}$ & ~$2.67\times10^{-1}$ & ~$1.94\times10^{-1}$ & ~$9.11\times10^{-2}$ & ~$0$ & ~$0$ & ~$0$ \\ 
\hline\hline
\end{tabular}
\label{table:P_extremal_KN}
\end{table}

\subsection{Nonextremal black holes}

\begin{figure}[t]
\centering
\includegraphics[width=15cm]{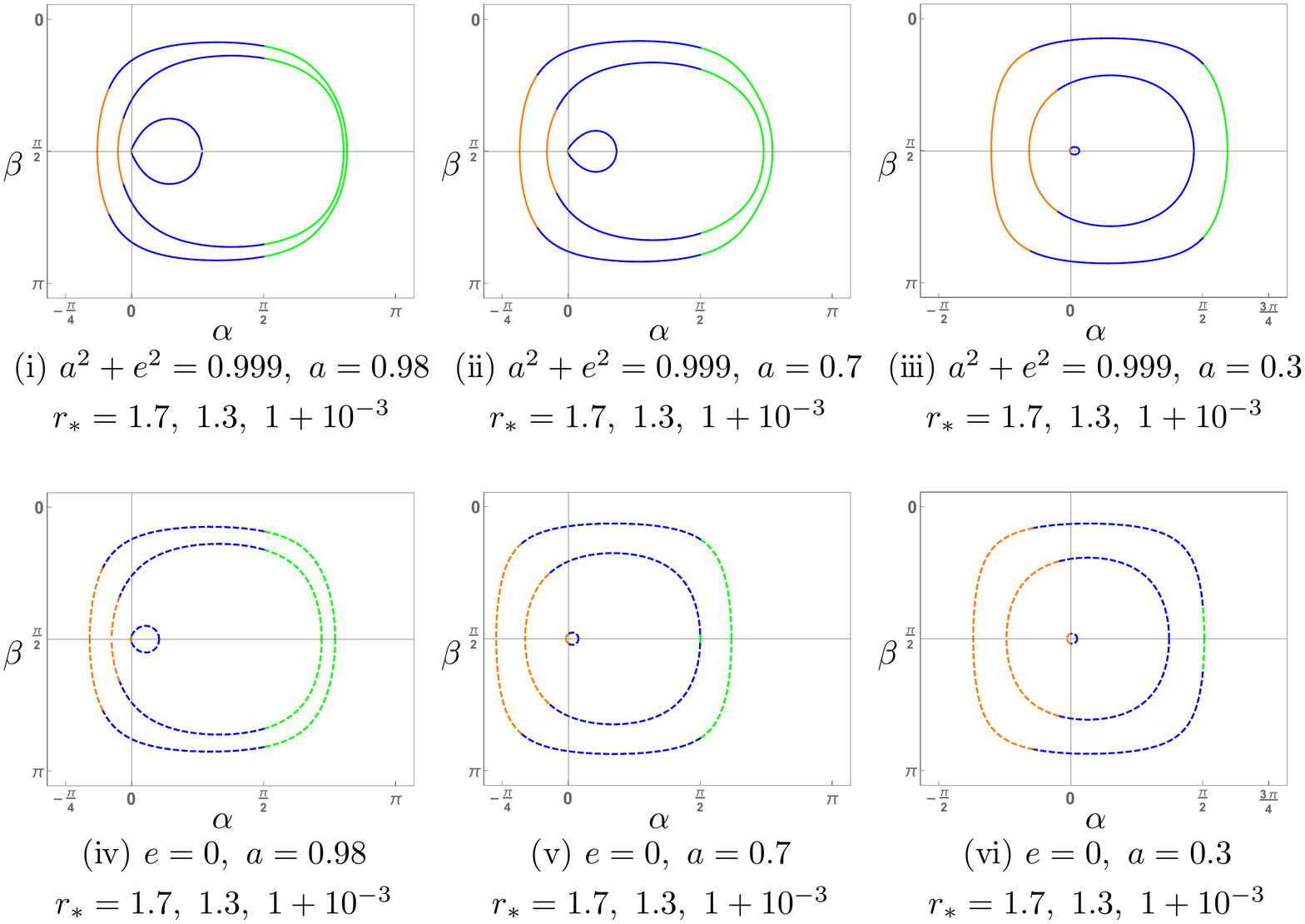}
\caption{
Critical angles in $\a$-$\b$ plane in nonextremal black holes.
The green, blue, and orange lines show ($\a_{\rm 1(b)},\b_{\rm 1(b)}$), ($\a_{\mathrm{1(c)}},\b_{\mathrm{1(c)}}$), and ($\a_2,\b_2$), respectively.
The solid lines (the upper three panels) and dashed lines (the lower three panels) are the cases of $a^2+e^2=0.999$ and $e=0$, respectively.
For each panel, we set $r_*=1.7$, $1.3$, $1+10^{-3}$ from outside to inside.
}
\label{fig:EC_KN}
\end{figure}

\begin{figure}[t]
\centering
\includegraphics[width=10cm]{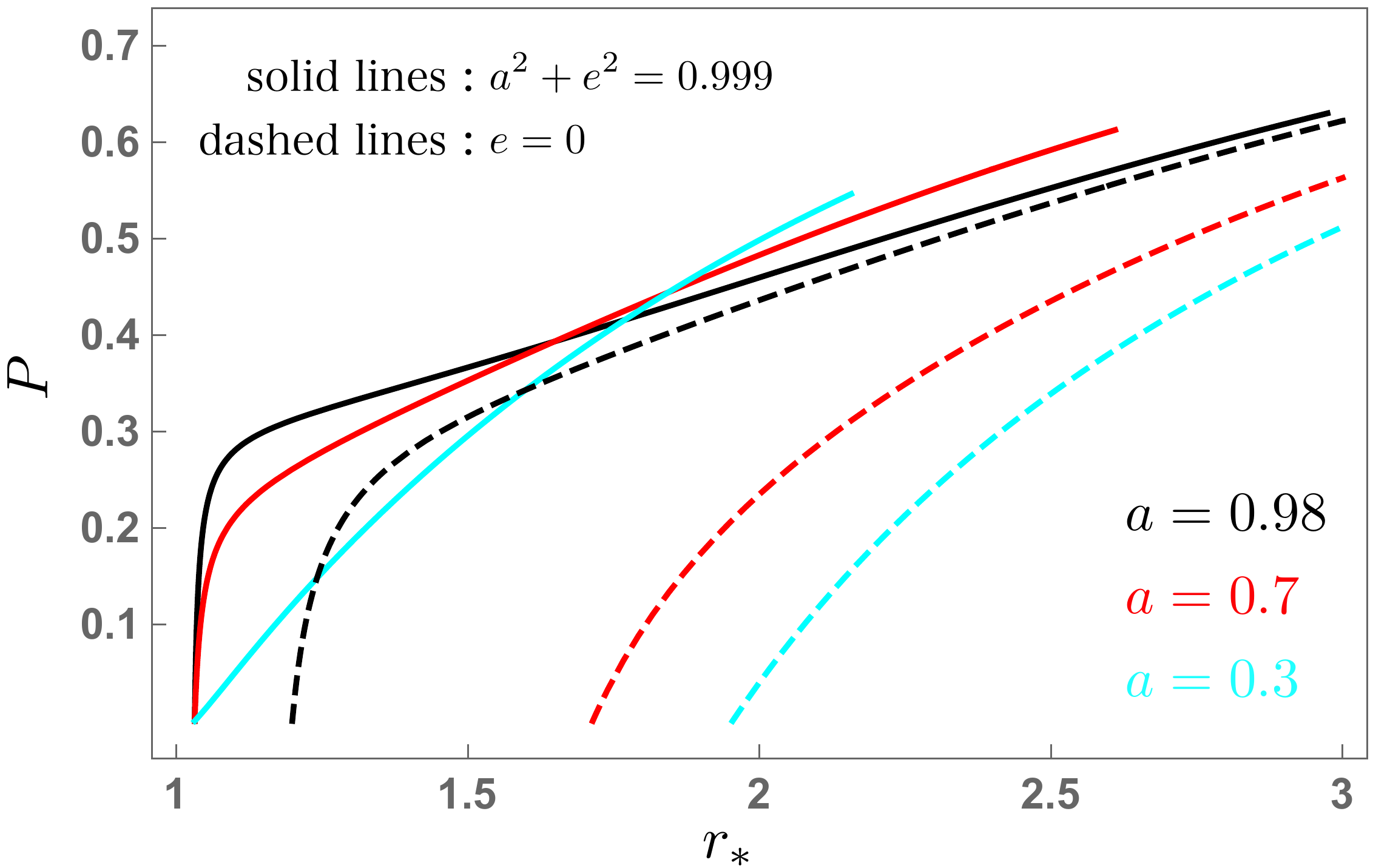}
\caption{
Escape probability in 
nonextremal black holes.
The solid and dashed lines are the cases 
of $a^2+e^2=0.999$ and $e=0$, respectively.
}
\label{fig:P_KN}
\end{figure}

Finally, we show some examples of the critical angles and escape probability in nonextremal black holes, where $a^2+e^2=0.999$ with $a=0.98,~0.7,~0.3$ and $e=0$ with $a=0.98,~0.7,~0.3$.
The critical angles and escape probability with the above cases are shown in Fig.~\ref{fig:EC_KN} and Fig.~\ref{fig:P_KN}, respectively.
We find that both the sizes of $S$ and $P$ are monotonically decreasing 
as $r_*$ decreases.
In the horizon limit, $S$ shrinks to the origin of the $\alpha$-$\beta$ plane and $P$ becomes zero.
However, in rapidly rotating and near-extremal cases, even if $r_*$ is close to $r_{\mathrm{H}}$, the escape cones have a relatively large size [see Fig.~\ref{fig:EC_KN}(i) and \ref{fig:EC_KN}(ii)].
Similarly, the escape probability takes a relatively large value up to the very vicinity of the horizon, and finally approaches zero rapidly.


\section{Conclusion and discussions}
\label{sec:6}

We have examined the escape of photons from the vicinity of the horizon to infinity in the Kerr--Newman black hole spacetime. We have evaluated escape cones at an emission point staying at rest in the LNRF and also the escape probability under assuming isotropic emission. Our main result is that when the black hole is extremal and has the spin parameter larger than 1/2, even if the light source arbitrarily approaches the horizon, the escape cone does not shrink and the escape probability remains nonzero.
Furthermore, the nonzero escape probability in the horizon limit monotonically increases as the spin increases from 1/2 with remaining extremal. Consequently, in the extremal Kerr black hole, it takes the largest value, $0.2916\cdots$.

The reason for the nonvanishing photon escape probability in the horizon limit is that
photons can be reflected at a radius arbitrarily close to the horizon.
This phenomenon is also related to the fact that the radius of the innermost spherical photon orbit coincides with the horizon radius.

Even for extremal black holes, the escape probability in the horizon limit becomes zero if the spin parameter is equal to or less than 1/2. Similarly, it does in the case of nonextremal black holes. However, in rapidly rotating and near-extremal black hole spacetime, as the emission point approaches the horizon, the escape probability keeps a relatively large nonzero value until it reaches a near-horizon region, and finally becomes zero when it reaches just outside the horizon. 
Even in this case, the fact that photons can be reflected at a radius very close to the horizon is essential~\cite{Igata:2019pgb}.

Our result is that up to approximately 30 percent of photons emitted from the light source near the horizon of the rapidly rotating (near-)extremal black hole escape to infinity. Hence, the phenomenon near the horizon of such a rapidly rotating and (near-)extremal black hole (e.g., the Penrose process or the black hole shadow) must be relatively visible compared to the case of a slowly rotating black hole. We speculate the effect of the nonzero escape probability on black hole shadow observations. Photons arriving at the shadow edge have been scattered at the radii of spherical photon orbits in the past. In particular, for rapidly rotating black holes, photons reaching the innermost shadow edge have been scattered at almost the horizon radius. Therefore, observing the innermost shadow edge is equivalent to observing a near-horizon region of the black hole. In fact, one of the estimates based on the observations of the M87 galactic center suggests that the central object may be rapidly rotating~\cite{Feng:2017vba}. Therefore, according to our results, a sufficient number of photons scattered at almost the horizon radius reach the innermost edge of the shadow, so that we can observe a near-horizon region through them.

\begin{acknowledgments}

The authors thank T.~Tanaka, K.~Yamada, T.~Narikawa, K.~Nakashi, and K.~Nakao for useful comments.
This work was supported by
JSPS KAKENHI Grants No.~18J10275~(K.O.), No.~JP19K14715~(T.I.), No.~JP19K03876~(T.H.), No.~JP18K03652~(U.M.) 
and the MEXT-Supported Program for the Strategic Research Foundation at Private Universities 2014--2017 (S1411024)~(T.I., T.H.).
K.O.~is grateful to the Yukawa Institute for Theoretical Physics at Kyoto University, where this work was developed during 
``The 3rd Workshop on Gravity and Cosmology by Young Researchers"(YITP-W-18-15).
\end{acknowledgments}


\end{document}